# Nanomechanical resonators show higher order nonlinearity at room temperature

Madhav Kumar[1], Bhaskar Choubey[2] and Harish Bhaskaran[1§]

1 Department of Materials, University of Oxford, United Kingdom
2 Department of Engineering Science, University of Oxford, United Kingdom
§[2]harish.bhaskaran@materials.ox.ac.uk, §madhav.kumar@materials.ox.ac.uk

Most mechanical resonators are treated as simple linear oscillators. Nonlinearity in the resonance behaviour of nanoelectromechanical systems (NEMS) has only lately attracted significant interest. Most recently, cubic-order nonlinearity has been used to explain anomalies in the resonance frequency behaviours in the frequency domain. Particularly, such nonlinearities were explained using cubic nonlinearity in the restoring force (Duffing nonlinearity) or damping (van der Pol nonlinearity). Understanding the limits of linear resonant behavior is particularly important in NEMS, as they are frequently studied for their potential in ultrasensitive sensing and detection, applications that most commonly assume a linear behavior to transduce motion into a detected signal. In this paper, we report that even at low excitation, cubic nonlinearity is insufficient to explain nonlinearity in graphene NEMS. Rather, we observe that higher order, in particular the fifth order effects need to be considered even for systems at room temperature with modest quality factors. These are particularly important results that could determine the limits of linear detection in such systems – and quite possibly present unconventional avenues for ultrasensitive detection paradigms using nonlinear dynamics. Such intriguing possibilities, however hinge crucially on a superior understanding and exploitation of these inherent nonlinearities as opposed to modelling them as approximated linear or cubic systems.

## Introduction

The majority of NEMS are designed as extensions of linear mechanical resonators. Such designs generally ignore any nonlinearity in the device parameters; more generally, systems are designed to operate in the linear regime[1-2]. Unfortunately, in NEMS, nonlinearity is triggered at very low oscillation amplitudes, thus restricting the linear regime of operation[3-4]. This is especially true for emerging NEMS based on graphene and other 2D materials[5-13] where recent studies have explored such nonlinearities both experimentally[6-11] as well as theoretically[14-15]. However, there has been limited quantification of such nonlinearity *vis-a-vis* the linear response. Knowledge of how this will affect the performance of graphene sensors is nascent. Thus, it is important to study all nonlinearities possible in such devices which would affect their operation as sensors. In this paper we explore the nature of higher order nonlinearities in such resonators. We perform experiments on a monolayer graphene resonator and show that there are significant challenges at the intersection of NEMS with inherent nonlinear dynamics, especially in regards to sensing applications. Underlining the complexity of the resonant behaviours, we show that quintic nonlinearity, an often-ignored higher order term, is crucial in describing their behaviour more accurately.

**Materials and Methods**

**Fabrication of Multiply Clamped H-shape Graphene Resonators**

We have fabricated a multiply clamped H-shape CVD grown graphene resonator by using standard nanofabrication processes reported previously[9]. The process starts with patterning H-shape graphene on $SiO_2$/Si wafer using electron beam lithography (EBL) and reactive ion etching (RIE). Gold electrodes are fabricated as electrical contact for the devices. EBL is used to pattern electrodes followed by thermal evaporation of 70 nm Au on 8 nm Cr (adhesive layer). Selective etching of underlying $SiO_2$ is performed using buffered solution of hydrofluoric acid (BOE) and devices are dried using critical point drying.

**Piezoresistive Readout of Graphene Resonators**

The measurement setup used here is similar to that shown previously[9,16]. Here we have employed piezomechanical excitation by applying voltage ($V_{Ext}$) to the piezo shaker. Hence by changing $V_{Ext}$ we change the amplitude of vibration of motion of graphene resonator. The mechanical vibration of the graphene resonator is measured piezoresistively[9]. The DC bias ($V_{Inp}$) also induces joule heating in the graphene resonator, hence varying $V_{Inp}$ allows us to vary the temperature of the resonator. All measurements are performed at room temperature and at pressures of $\sim 4.4 \times 10^{-6}$ Torr.

**Results**

**Extraordinary Resonance Shape Observed in Graphene NEMS**

To study such nonlinearity, we consider an H-shaped, piezomechanically excited and piezoresistively measured resonators[9,16]. Figure 1a shows an SEM micrograph (with false colour) of one such resonator with a circuit diagram of the measurement setup. A measured frequency response at constant DC power supply to the device ($V_{Inp}$) and with varying external drive applied to the piezo shaker ($V_{Ext}$) is shown in Figure 1b. The piezo shaker serves to drive the motion of the device around the desired frequency of interest. Resonance frequency of the device is measured at around 1.190 MHz. Interestingly, we observe that as we increase the external drive voltage ($V_{Ext}$), the shape of the resonance frequency response changes extraordinarily .

Such a frequency response suggests that at $V_{Ext}$ = 2 V the resonator exhibits a weak Duffing nonlinearity[4], but as we increase the drive amplitude to $V_{Ext}$ ≥ 3V we start to observe a completely different shape of the resonance frequency response. This response not only displays a sharp resonance peak at 1.1908 MHz, but also shows other resonance peaks at 1.1928 MHz.

**Device modeled as a cubic nonlinear resonator**

Nonlinearity has been previously explained using analytical expressions of the equation of motion of a resonator. Most mechanical resonators have a nonlinear restoring force acting on them at large amplitudes of motion[3-4,6,17]. This leads to their motion becoming that of a higher order nonlinear system[4,6]. Such higher order nonlinear systems have previously been modelled by taking into account the cubic nonlinearity in spring constant (Duffing) and damping (van der Pol) as shown in Equation 1[4, 6].

Here, $M$ is the effective mass of the resonator, $k$ is the linear spring constant, $x$ is the displacement, $F$ is the amplitude of the driving force and $\omega$ is the frequency of oscillation. $\alpha$ is the cubic order non-linearity of the spring constant, often referred to as the Duffing parameter. It can be negative as well as positive depending on whether the resonator is 'softening' or 'hardening' as we increase the amplitude of

$$M\left(\frac{d^2x}{dt^2}\right) + (\gamma + \eta x^2)\left(\frac{dx}{dt}\right) + kx + \alpha x^3 = F\cos(\omega t) \quad \text{... Equation 1}$$

excitation[4]. $\gamma$ is the linear and $\eta$ the first order nonlinear coefficients appearing in the damping terms for most resonating systems.

Initially, we attempted to numerically fit Equation 1 to the frequency response data of the device. Figure 2 shows this numerical fit of equation 1 considering cubic nonlinearity for the device at $V_{Ext}$ = 4V. We found that our experimental data departs from classical Duffing behavior. This indicates that the nonlinear response of these single layer graphene resonators with high sensitivity at higher drive amplitudes cannot be fully explained by using Equation 1 (cubic nonlinearity in spring and damping). This was initially puzzling, especially as the graphene resonators studied in this report were measured at room temperature, thus potentially having less sensitivity than previous studies of nonlinearity at cryogenic temperatures. One hypothesis was that there might be an influence of even higher order nonlinearities on this behavior, especially as systems have previously shown to have ultrasensitivity[9].

**Quintic nonlinearity: The 5$^{th}$ order nonlinearity in Resonator**

To test the hypothesis of even higher order nonlinearities, and also to obtain a limit of the order of nonlinearity that could potentially play a role in the nature of the resonance, we looked at the higher harmonics of oscillation of the resonator. This is because, for nonlinear systems the frequency response due to harmonic force excitation at *f* will have sub harmonics (*f* = n$f_0$), where n ≠ 1[3,17-18].

Figure 3a shows a representation of a resonator with linear as well as higher order nonlinear spring constants (k, α, β). While driving the device at its resonance frequency $f_o$ we measured the amplitude of resonance simultaneously not only at $f_o$ but also at higher harmonics as shown in Figure 3b. Observation of sub-harmonics at around 3$f_o$ along with the main resonance peak at $f_o$ in higher order nonlinear systems has been predicted theoretically[3,17-18] as well as demonstrated experimentally[7-9]. However, observation of sub harmonics at 5$f_o$ has not been observed in such a system. This therefore indicates that our system is a quintic (5$^{th}$ order) nonlinear resonator, confirming our initial hypothesis.

Further as we increase the external drive at resonance frequency (V$_{Ext}$ ($f_0$)), the amplitude of vibration not only increases at $f_0$ but also at *3$f_0$* and *5$f_0$*. This serves to further confirm that the three modes 5$f_o$, 3$f_o$ and $f_o$ are coupled to one another and have an effect on each other.

**Confirmation of 5$^{th}$ order nonlinearity in Graphene Resonator**

Generally, due to extremely low values of higher order restoring terms, they are neglected in the equation of motion for NEMS resonators By confirming our original hypothesis that the fifth order nonlinearity β along with α might be required to fully explain the motion of higher order nonlinear systems, we proceeded to model such a system which can be written as Equation 2.

$$M\left(\frac{d^2x}{dt^2}\right) + (\gamma + \eta x^2)\left(\frac{dx}{dt}\right) + kx + \alpha x^3 + \beta x^5 = F\cos(\omega t) \quad \text{... Equation 2}$$

Figure 4a is the numerical solution of equation 2, for normalized mass and particular values of F. Figure 4b shows the experimental resonance curve of the device at V$_{Ext}$ = 4 V. Comparing the numerical and experimental data in Figure 4, we observe that the particular shape of resonance curve of the device at higher drive amplitudes can be explained far better using equation 2 rather than equation 1, thereby confirming higher order nonlinearity, in particular the fifth order nonlinearity in our device, as indicated by sub harmonics shown in Figure 3. This is evidence that the resonator is so sensitive that higher order nonlinearity can no longer be neglected and has to be considered along with the Duffing nonlinearity. Thus Equation 2 explains the jump in the amplitude (saddle node bifurcation) with an additional resonance peak of relatively lower amplitude at $V_{Ext}$ = 4V (Figure 4b). Indeed, equation 2 explains the unique shape of the observed frequency response at high driving amplitudes confirming the presence of higher order nonlinearity at higher amplitudes of motion. Furthermore, by comparing the change of phase data of the device with the numerical solution of equation 2 confirms the change in phase in both experimental as well as numerical data.

**Alternative manifestations of quintic nonlinearity and their numerical predictions:**

In Figure 5a we recall Figure 1b to compare the experimental data to the numerical solution of equation 2 which includes the $5^{th}$ order nonlinearity. Figure 5b shows this solution with the same coefficients used previously and varying the amplitude of drive F. It is seen that until lower value of F the resonance characteristics resembles the Duffing nonlinearity. However, we observe a unique change in the shape of the resonance curve as we increase F to higher values, which resembles the experimental data shown in Figure 5a. Increasing F further results in following the usual mechanical resonator behavior i.e. amplitude of the resonance frequency response increases while the resonance frequency decreases. We observe a similar behavior in the frequency response of our device with changing drive amplitude $V_{Ext}$ (F in the case of numerical data). As explained earlier, at $V_{Ext}$ = 2 V the resonator behaves more like a weak Duffing system nonlinearity, and upon increasing $V_{Ext} \geq$ 3V it resembles the numerical solution response for F $\geq$ 1.7 (Figure 5b). It is worth mentioning that in both experimental data and numerical solution we just vary the driving amplitude while keeping all the other parameters constant. This confirms that our system shows the presence of quintic nonlinearity.

**No extra modes due to particular shape of resonator**

To confirm that no extra modes exist close to first few modes due to the particular shape of resonator, we performed modal analysis using *COMSOL Multiphysics*. Figure 6 shows that the mode shape of four side clamped beam resonator are same as that of double side clamped rectangular shape beam resonator and no extra modes exists for multiple clamped H shape resonator. Hence, we rule out the fact that the different modes could generate extra peak at high amplitude of motion and interfere with the main resonance peak.

Kumar *et al.***Discussion and Conclusions**

Our results thus demonstrate that even at room temperature, given the very small masses of these mechanical resonators and their intrinsically high sensitivity, much higher order effects such as the quintic restoring nonlinearity play a crucial role. Further, we have observed subharmonics along with the natural resonance confirming that indeed these higher order modes are not just of scientific interest but also of interest in the design and operation of such devices. Such subharmonics could be used to operate and sense at much higher frequencies than those permitted by mechanical resonators, introducing an order of magnitude improvement in sensitivity. These sub (cubic and quintic) harmonics in graphene resonators, which are coupled to one another and with first harmonics, further confirms that the higher order nonlinear terms affect the resonance frequency response in graphene NEMS at even modest amplitudes of motion. Importantly, we observe quintic nonlinearity at room temperature, thus ruling out that such effects are limited to low temperature experiments.

As we have shown here, modeling this nonlinearity even using a cubic-order model would lead to completely erroneous estimates of sensitivity. Now if we consider that sensor designs today use linear paradigms in their design philosophy, the operation regime of NEMS governed by higher order nonlinearity, resonance peak splitting, mode coupling, nonlinear damping and other nonlinearities, present even greater challenges. NEMS using graphene and other 2D materials (and indeed, more generally other highly sensitive NEMS) are inherently nonlinear, to higher order. If one cannot avoid operation in the nonlinear regime (as we believe is the case), a superior understanding of this nonlinearity is crucial for future designs of such sensors.

**Contributions**

MK performed the modeling, fabrication, experimentation, data acquisition and analysis. HB conceived of the concept and contributed to the experimentation. BC conceived of the concept and contributed to the study of non-linear response in resonators. All three authors co-wrote the manuscript.

**Acknowledgments**
**Acknowledgments**

We knowledge support from the OUP John Fell Fund and EPSRC via grants EP/J00541X/1/2 and EP/J018694/1.

**Conflict of Interests**

The authors declare no competing financial interest.

Kumar *et al.*

Kumar *et al.*

**Figures**

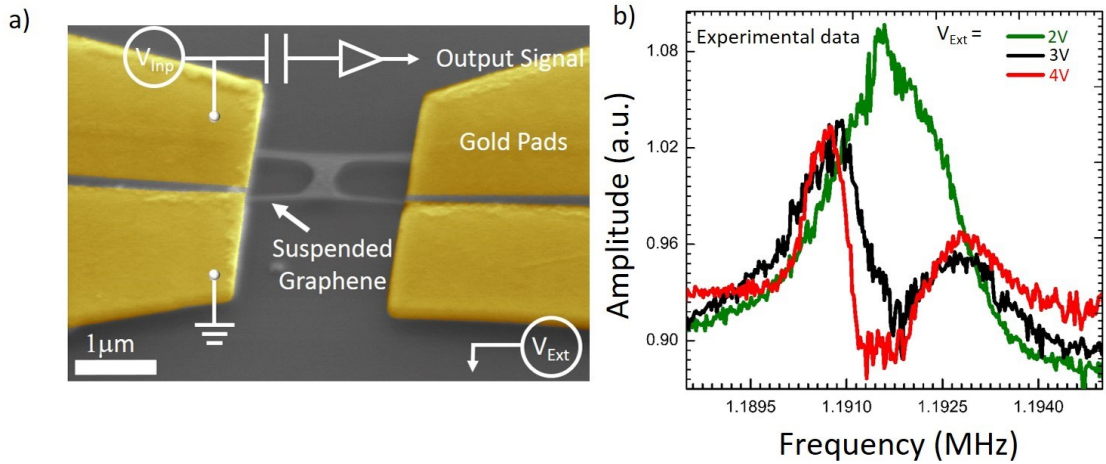

**Figure 1.** Nonlinearity in graphene NEMS **a)** SEM micrograph of the device with a circuit diagram of the measurement setup. **b)** Resonance frequency response ($f_0$) of the device with varying $V_{Ext}$ and constant $V_{Int}$.

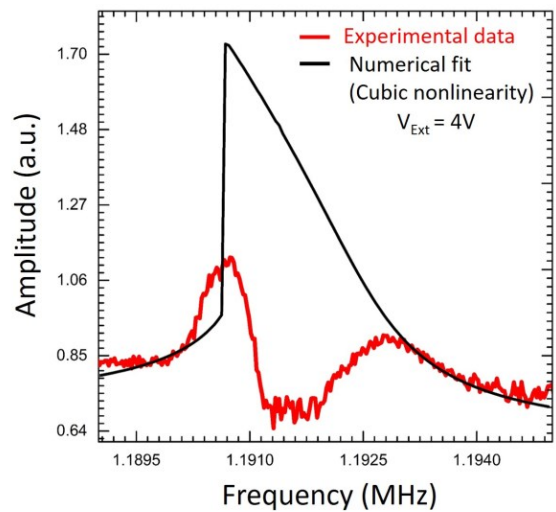

**Figure 2.** Device modeled as a cubic nonlinear resonator: Resonance frequency response ($f_0$) at $V_{Ext}$ = 4V for device (red line). Numerical data (solution of equation 1) fitted to the experimental data for the device (black line). The cubic solution is not a fit to the experimental data.

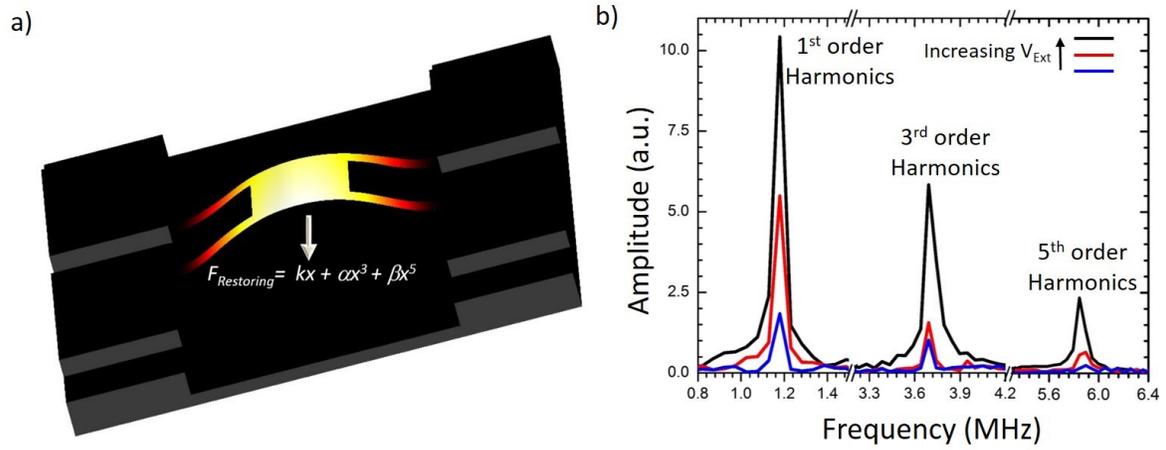

**Figure 3.** Quintic nonlinearity. **a)** Schematic of resonator with linear as well as nonlinear spring constants (k,α,β). **b)** Observation of sub-harmonics. Device showing 1st ($f_0$), 3rd and 5th harmonics (sub-harmonics) using external excitation at $f_0$ for different excitations ($V_{Ext}$). This close correlation indicates 5th order nonlinearity in the spring constant.

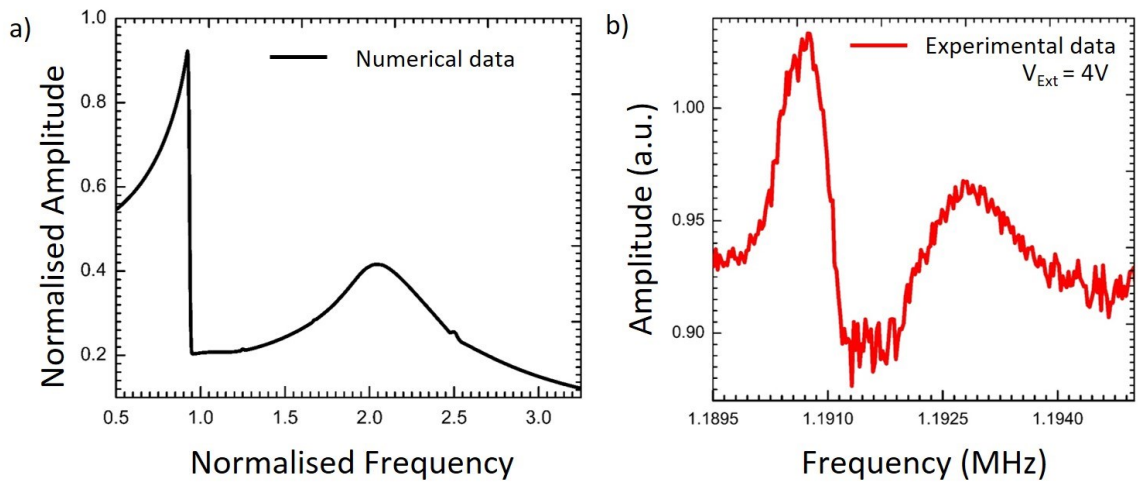

**Figure 4.** Confirmation of 5th order nonlinearity. **a)** Numerical solution of equation 2 **b)** Experimental data of device for $V_{Ext}$ = 4V. This close correlation indicates 5th order nonlinearity in spring constant of the device.

Kumar *et al.*

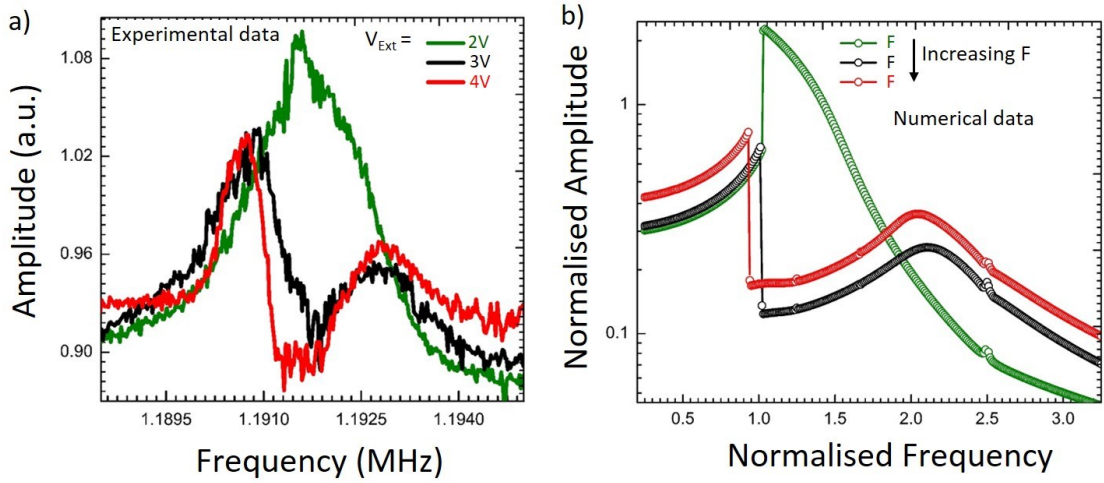

**Figure 5.** Alternative manifestations of quintic nonlinearity and their numerical predictions (**a**) Variation of amplitude at $f_o$ of device with varying $V_{Ext}$ and constant $V_{Int}$. (**b**) Numerical solution of equation 2 for different F (solution normalized).

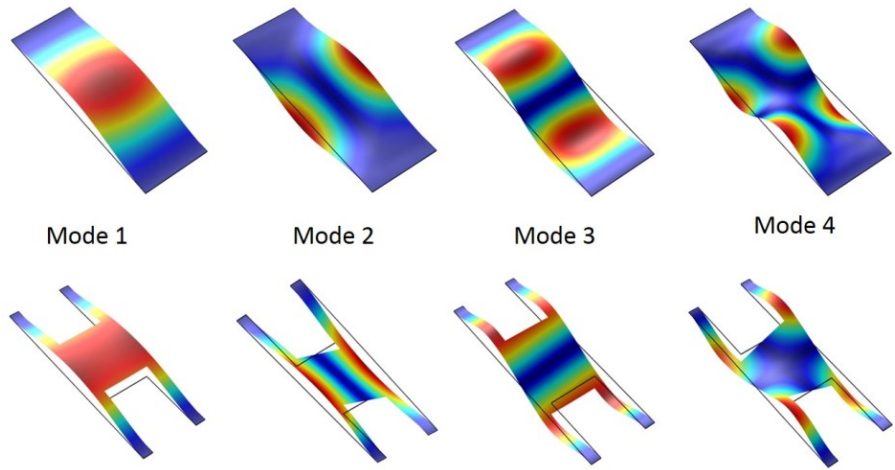

**Figure 6** : First four mode shape of beam resonator for a) double side clamped rectangular (top) and b) four side clamped (bottom); calculated using finite element modeling.